\documentclass[11pt]{article}
\usepackage{epsfig,dsfont}

\begin{document}
\sffamily

\thispagestyle{empty}
\vspace*{15mm}

\begin{center}

{\LARGE Static quark-antiquark potential and 
\vskip2mm
Dirac eigenvector correlators}

\vskip20mm
Erek Bilgici and Christof Gattringer
\vskip5mm
Institut f\"ur Physik, Unversit\"at Graz \\
8010 Graz, Austria 
\end{center}
\vskip30mm

\begin{abstract}
We represent the Polyakov loop correlator as a spectral sum of
correlators of eigenvectors of the lattice Dirac operator. 
This spectral representation is studied numerically using quenched 
SU(3) configurations below and above the deconfinement temperature. 
We analyze whether the individual Dirac eigenvector correlators differ 
in the confined and deconfined phases. The decay properties of the 
normalized Dirac eigenvector correlators turn out to be essentially 
identical in the two phases, but the amplitudes change. This change 
of the amplitudes shifts the relative contributions of the individual 
Dirac eigenvector correlators and is the driving mechanism for
the transition from the confining static potential into the deconfining one. 
\end{abstract}

\setcounter{page}0
\newpage
\noindent
{\Large  Introductory remarks}
\vskip3mm
\noindent
Confinement of quarks is one of the key features of QCD. Of particular interest
is the fact that at a critical temperature QCD undergoes a transition
into  a deconfined high temperature phase. Understanding the transition and
characterizing  QCD in the deconfined phase are widely discussed problems. 

It is obvious, that confinement and deconfinement, 
respectively, must be manifest
in the Dirac operator and its inverse, the quark propagator, since the quarks
need to know whether they are confined or not. In a series of recent papers 
\cite{paper1} -- \cite{wipf2} 
the connection of quantities
related to confinement, e.g., the expectation value of the Polyakov loop, to
spectral sums of the Dirac operator was studied. Signatures of confinement in
the Dirac spectrum and their change at the QCD transition were identified.

In this article we go beyond results for only the Dirac eigenvalues and study a
spectral representation of the static quark anti-quark potential. Since this
quantity depends on the relative spatial position of the static sources, its
spectral representation contains also the eigenvectors of the Dirac operator. In
particular the shape of the potential must reflect itself in spatial
correlators of Dirac eigenvectors. Very little is known about such correlators
of Dirac eigenvectors and we study these relations numerically using
quenched lattice QCD configurations. We explore how the spectral signatures of
Dirac eigenvalues and eigenvectors change at the deconfinement transition.   

We  remark at this point that very recently Synatschke, Wipf and Langfeld
have studied correlators of spectral sums with a different, 
IR dominated spectral function \cite{wipf2}. 
Using the boundary condition technique of 
\cite{paper1} they project to loops with the same transformation properties 
as the Polyakov loops which we use here. The continuum arguments given in 
\cite{wipf2} and the numerical study with SU(2) gauge configurations provide
interesting complementary insight to the analysis presented here.

\vskip6mm
\noindent
{\Large Dirac operator and Polyakov loops}
\vskip3mm
\noindent 
In this section we repeat and refine the arguments \cite{paper1} 
which lead to the connection of the Polyakov loop and its correlators to
spectral sums for the Dirac operator. 

The Polyakov loop at spatial position $\vec{x}$ is given by
\begin{equation}
L(\vec{x}) \; = \; \frac{1}{n} \, \mbox{tr}_c  \left[ \,
\prod_{t=1}^{N}  U_4(\vec{x},t) \right] \; .
\end{equation}
$U_4(\vec{x},t)$ is the temporal component (= 4-component) of the gauge
variables $U_\mu(\vec{x},t) \in$ SU($n$) and tr$_c$ is the color trace. The rank
$n$ of the gauge group in the fundamental representation 
appears in the overall normalization of the Polyakov
loop. The product runs over all $N$ temporal links at a given spatial point
$\vec{x}$. We work on a lattice of size $L^3 \times N$ and  for the gauge field
use boundary conditions which are periodic in all four directions. We are
interested in correlators of the Polyakov loop 
\begin{equation}
\langle L(\vec{x}) L^*(\vec{y}) \rangle \; \sim \;
A \, e^{-N \, V(|\vec{x} - \vec{y}|)} \; ,  
\label{loopcorr}
\end{equation}
where $^*$ denotes complex conjugation and
$V(|\vec{x} - \vec{y}|)$ is the static quark-antiquark potential.

The staggered Dirac operator at vanishing quark mass is given by
(we set the lattice spacing to 1) 
\begin{equation}
D(x,y) = \frac{1}{2}
\sum_{\mu=1}^4 \eta_\mu(x) \, \Big[ 
U_\mu(x) \, \delta_{x+\hat{\mu},y} \, - \, U_\mu(x-\hat{\mu})^\dagger \,
\delta_{x-\hat{\mu},y} \Big] ,
\label{stagdir}
\end{equation}
where $x$ and $y$ are integer valued 4-vectors labeling the lattice sites
and $\eta_\mu(x) = (-1)^{x_1 + \, ... \, + x_{\mu-1}}$ is the staggered sign
function. 

The terms of the staggered Dirac operator (\ref{stagdir}) 
connect nearest neighbors. When powers of $D$ are considered, these terms
combine to chains of hops on the lattice. Along these chains 
products of the link variables $U_\mu(x)$ 
are collected. Taking the $m$-th power will give rise to chains 
of length $m$.
We now consider the $N$-th power of $D$, where $N$ is the 
temporal extent of our lattice. Thus we will encounter chains with a
length of $N$. Furthermore we set the two space-time arguments
of $D$ to the same value, $y = x$, such that we pick up only closed chains, 
i.e., loops starting and ending at $x$. Among these are the loops where 
only hops in time direction occur such that they close around compact
time. We obtain 
\begin{eqnarray}
\noindent
\mbox{tr}_c \big[ D^{N}(\vec{x},t|\vec{x},t) \big] & = & \mbox{trivial loops} 
\, + \,
\frac{(-1)^{N(x_1+x_2+x_3)}}{2^N} \, 
\mbox{tr}_c \prod_{s=1}^{N} U_4(\vec{x},s) 
\nonumber
\\
&& \hspace{20.5mm} + \;
\frac{(-1)^{N(x_1+x_2+x_3)}}{2^N} \, 
\mbox{tr}_c \prod_{s=0}^{N-1} U_4(\vec{x},N\!-\!s)^\dagger
\nonumber
\\
& = & \mbox{trivial loops} \, + \, 
\frac{1}{2^N} \, n L(\vec{x}) \; + \; 
\frac{1}{2^N} \, n L^*(\vec{x})\; .
\label{dandloops}
\end{eqnarray} 
The factor $(-1)^{N(x_1+x_2+x_3)}$ comes from the products of the staggered sign
factors $\eta_4(x)$. Here we use even $N$, such that this factor is equal to
$+1$. We stress that the forward and backward running Polyakov loops are the
only ones that wind non-trivially around compact time.

We now explore the fact that the Polyakov loops respond differently to a change
of the boundary conditions compared to other, trivial (= non-winding) loops. We
can change the temporal boundary condition of the Dirac operator by multiplying
all temporal link variables at $t = N$ with some phase factor  $z \in
\mathds{C}, |z| = 1$: $U_4(\vec{x}, N) \rightarrow \;  z \, U_4(\vec{x}, N) \;
\forall \, \vec{x}$. We evaluate the left-hand side of (\ref{dandloops}) using 
the Dirac operator in the transformed field, and denote the corresponding  Dirac
operator as $D_r$, where $r$ labels the boundary  conditions we use. Actually we
linearly  combine three different boundary conditions $z_r$, $r = 1,2,3$ to
obtain:
\begin{eqnarray}
&& \frac{2^N}{n} \sum_{r=1}^3 
c_r \, \mbox{tr}_{c} \big[ D_{r}^{N}(\vec{x},t|\vec{x},t) \big] 
\, = \, 
L(\vec{x}) \times \sum_{r=1}^3 c_r \, z_r \, 
\nonumber 
\\
&& + \; L^*(\vec{x}) \times \sum_{r=1}^3 c_r \, z_r^*
\; + \; \mbox{trivial loops} \times \sum_{r=1}^3 c_r \, .
\end{eqnarray}
The complex coefficients $c_r$  are determined by the requirement that the
factors in front of the complex conjugate loop $L(\vec{x})^*$ and in front of
the trivial  loops vanish, while the factor in front of the Polyakov loop
$L(\vec{x})$ is  required to be equal to 1. Here we use the following values for
the  $z_r$ and  the $c_r$,
\begin{equation}
z_1 = 1 \; , \; z_2 = e^{i 2\pi/3} \; , \; z_3 = e^{- i 2\pi/3} \quad 
\mbox{and} \quad 
c_r = z_r^* / 3 \; ,
\label{boundaryphase}
\end{equation}
which fulfill the above requirements. We conclude
\begin{equation}
L(\vec{x}) \; = \; \frac{2^N}{n N} \sum_{t=1}^N \sum_{r=1}^3 
z_r^* \, \mbox{tr}_{c} \, \Big[ D_r^{N}(\vec{x},t|\vec{x},t) \Big] \; .
\label{loopdirac}
\end{equation}
We have used the fact that the left-hand side is independent of $t$ and 
averaged the right-hand side over $t$. 

\vskip6mm
\noindent
{\Large Spectral sum for the Polyakov loop correlator}
\vskip3mm
\noindent
We now express the $N$-th power of the Dirac operator as a spectral sum. 
Let $\vec{v}^{\,(k)} (\vec{x},t)$ 
be the eigenvector of $D$ with eigenvalue $\lambda^{(k)}$. 
Then
\begin{equation}
D^N(\vec{x},t|\vec{x},t)_{c,c^\prime} = 
\sum_k \left(\!\lambda^{(k)}\!\right)^N 
\vec{v}^{\,(k)}(\vec{x},t)_c \, \vec{v}^{\,(k)}(\vec{x},t)_{c^\prime}^* \; , 
\end{equation}
where we have made explicit also the color indices $c, c^\prime$. 
The index $k$ runs over all eigenvalues of the Dirac operator. Inserting the
spectral sum into (\ref{loopdirac}) we find
\begin{eqnarray}
L(\vec{x}) & \, = \, & \frac{2^N}{n} \sum_k \sum_{r=1}^3 
z_r^* \,  \left(\lambda^{(k)}_r\right)^N \, 
\rho^{(k)}_r(\vec{x}) \; ,
\label{specsum} 
\\
\rho^{(k)}_r(\vec{x}) & \, = \, & \frac{1}{N} \sum_{t=1}^N \sum_{c=1}^n
\Big| \, \vec{v}^{\,(k)}_r(\vec{x},t)_{c} \, \Big|^2 \; ,
\label{density}
\end{eqnarray}
where again $r$ labels the three different boundary conditions we use and the 
index $k$ runs over all eigenvalues.

The spectral sum may be simplified further, using the fact
that the eigenvalues of the staggered Dirac operator are purely imaginary and
come in complex conjugate pairs. The eigenvectors corresponding to these pairs
are related to each other by multiplication with the staggered sign  $(-1)^{x_1
+ x_2 + x_3 + x_4}$, which leaves the densities (\ref{density}) invariant. Thus
in the spectral formula (\ref{specsum}) one needs to sum over only the
eigenvalues with positive imaginary part (we indicate this with the
superscript $+$
attached to the summation symbol)  and an extra factor of 2 appears. Thus we
obtain for the Polyakov loop correlator
\begin{equation}
L(\vec{x}) \, L^*(\vec{y}) \; = \;
\frac{4^{N + 1}}{9 n^2} \, {\sum_{k,l}}^{+} \! \sum_{r,s=1}^3 \!
z_r^* z_s 
\left(\!\lambda^{(k)}_r \lambda^{(l)}_s \!\right)^N  \! 
\rho^{(k)}_r(\vec{x}) \, \rho^{(l)}_s(\vec{y}) .
\label{speccorr} 
\end{equation} 
We stress that the result (\ref{speccorr}) is an exact formula, in particular it
holds for an individual gauge field configuration and not only in the ensemble
average. Furthermore, all terms involved in our spectral representation are
manifestly gauge invariant quantities.

Let us finally discuss the correlator after the vacuum expectation value  
$\langle .. \rangle$ has been evaluated. In this case one can use the fact that
the expectation value is invariant under spatial reflections, i.e., we can
interchange $\vec{x}$ and $\vec{y}$. When using the choice 
(\ref{boundaryphase}) for the phases at the boundary one obtains a particularly
simple form of the correlator
\begin{eqnarray}
&&\hspace*{-5mm} 
\Big\langle L(\vec{x}) L^*(\vec{y}) \Big\rangle \; = \;
\frac{4^{N + 1}}{9 n^2} {\sum_{k,l}}^{+} \!
\left[ \, 
\sum_{r=1}^3 \left\langle \,
\left(\!\lambda^{(k)}_r \lambda^{(l)}_r \! \right)^N 
\! \! \rho^{(k)}_r(\vec{x}) \, \rho^{(l)}_r(\vec{y}) \right\rangle \right.
\label{speccorrexp}  \\
&&
\hspace{52mm}  - \, \left.
\sum_{r<s} 
\left\langle \left(\!\lambda^{(k)}_r \lambda^{(l)}_s \! \right)^N \! \! 
\rho^{(k)}_r(\vec{x}) \, \rho^{(l)}_s(\vec{y}) \right\rangle \right] \; .
\nonumber
\end{eqnarray}
The final result is a sum of 6 terms, where the 3 correlators of the 
densities with equal boundary condition come with a plus sign, while 
the correlators with mixed boundary conditions are subtracted. 

\vskip6mm
\noindent
{\Large Truncated spectral sums}
\vskip3mm
\noindent
The formulas (\ref{speccorr}) and (\ref{speccorrexp}) constitute the announced
spectral representation of the Polyakov loop correlator and thus, via
Eq.~(\ref{loopcorr}), the spectral decomposition of the static quark-antiquark
potential. The right-hand side is a double sum over all Dirac eigenvalues and we
can indeed decompose the correlator into IR and UV parts by taking into account
only parts of the spectrum. We now analyze numerically how the individual terms 
in (\ref{speccorr}) and (\ref{speccorrexp}) give rise to the static quark
potential  and how they see the QCD phase transition. 

The numerical analysis is done on quenched SU(3) gauge configurations generated
with the L\"uscher-Weisz action \cite{LuWe} on various $L^3 \times N$ lattices
with $L = 10$ and 12 and $N = 4$ and 6. All data we show here are for the
largest lattice, $12^3 \times 6$, and we have a statistics of 20 configurations
both in the confined and the deconfined phase. These results were
cross-checked on smaller lattices where larger statistics of ${\cal O}(100)$
configurations were produced. The lattice spacing was
determined in \cite{scale} using a Sommer parameter of $r_0 = 0.5$ fm. The gauge
coupling was chosen such that we have ensembles in the confined phase at a
temperature of $T = 236$ MeV and ensembles in the deconfined phase at $T = 412$
MeV. Above $T_c$ ($\sim$ 300 MeV)
the theory has a non-vanishing Polyakov loop, which  
can have three possible phases.  In our numerical analysis we restrict
ourselves to configurations with (essentially) real Polyakov loop, since the
results for ensembles with one of the two complex phases of the Polyakov loop
may be related to the real case by a simple $\mathds{Z}_3$ transformation. For
the numerical analysis of the spectral sums  (\ref{speccorr}) and
(\ref{speccorrexp}) we computed complete sets of eigenvalues and eigenvectors of
the staggered Dirac operator using ScaLAPACK routines.
The results from the  spectral sums were cross-checked to an evaluation of the
Polyakov-loop correlators  directly from the gauge links and agreement of the
first 7 to 10 digits was found. 

The central quantity we consider in this section is the correlator $\langle
L(\vec{x}) L^*(\vec{y}) \rangle$, evaluated with the exact formula 
(\ref{speccorr}). In order to improve the statistics we average over $\vec{x}$
and $\vec{y}$, keeping the distance $d$ fixed. We define
\begin{equation}
C(d) \; = \; \frac{1}{3 L^3} \sum_{\vec{x}} \sum_{j = 1,2,3} 
\Big\langle L(\vec{x}) L^*(\vec{x}+d\,\widehat{e}_j) \Big\rangle \; ,
\end{equation}
where $\widehat{e}_j, j = 1,2,3$ are the unit vectors in the three spatial
directions and for $L(\vec{x}) L^*(\vec{x} + d\,\widehat{e}_j)$ we insert the
spectral representation (\ref{speccorr}). 
According to Eq.~(\ref{loopcorr}) $C(d)$ behaves as
\begin{equation}
C(d) \; \sim \;
A \, e^{-N \, V(d)} \; .  
\label{loopcorr2}
\end{equation}
We now study the correlator $C(d)$ but sum the spectral sums in 
(\ref{speccorr}) only up to  some
largest eigenvalue $\lambda_{cut}$. Thus for small $\lambda_{cut}$  we only see
the IR contributions and as we increase $\lambda_{cut}$ towards the largest
eigenvalue $\lambda_{max}$, we gradually recover
the full result. These truncated sums allow one to study how the full
correlator builds up from the different contributions in the IR and UV. 

\begin{figure}[t]
\begin{center}
\hspace*{-4mm}
\includegraphics[height=69mm,clip]{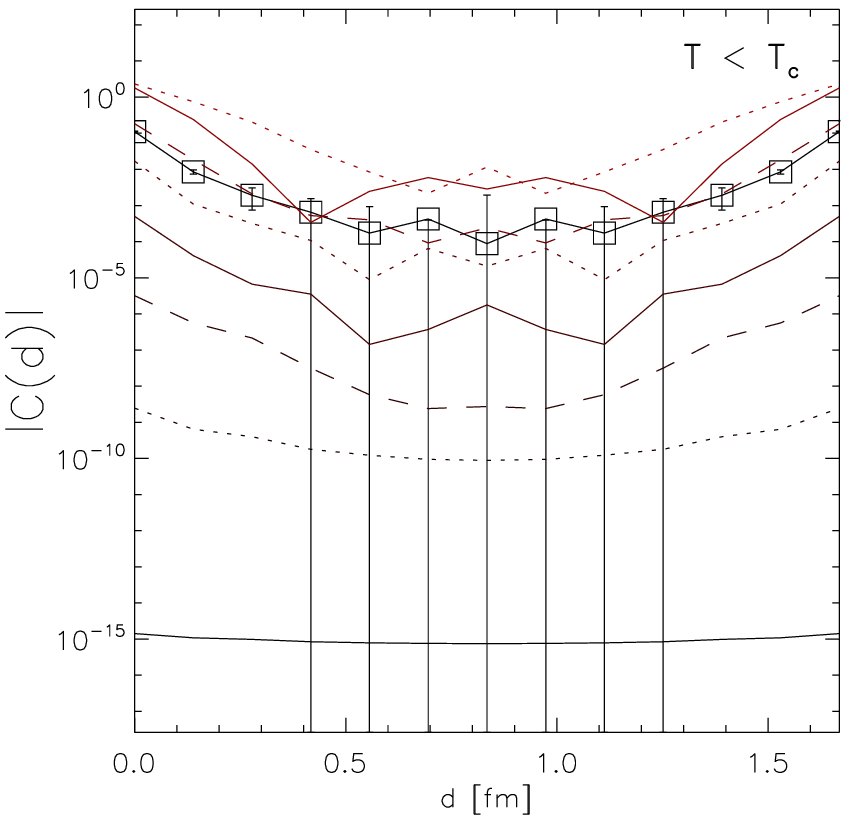}
\hspace{-5mm}
\includegraphics[height=69mm,clip]{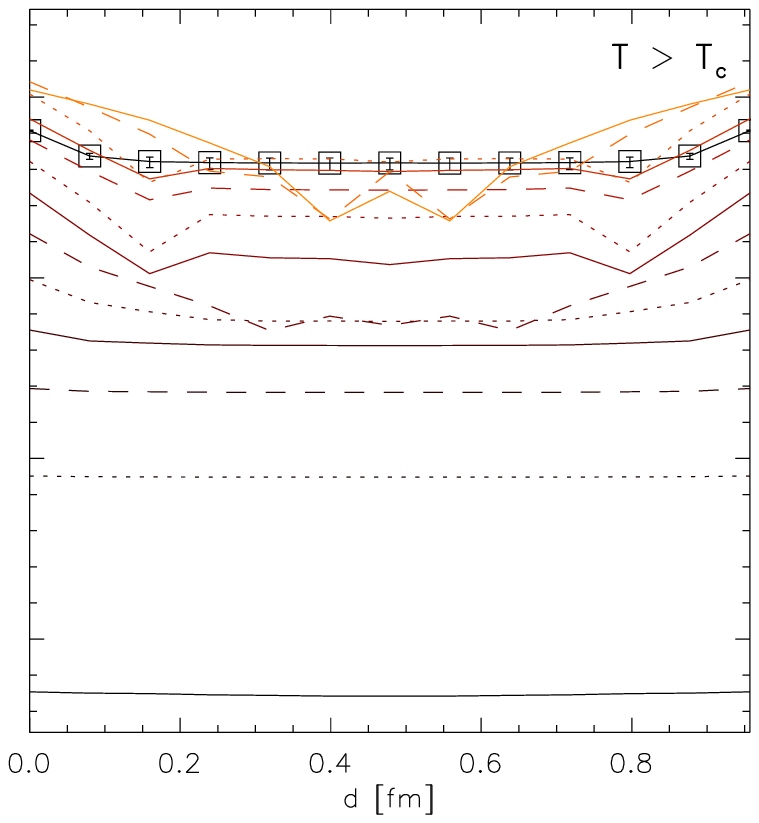}

\vskip4mm

\includegraphics[height=4.7mm,clip]{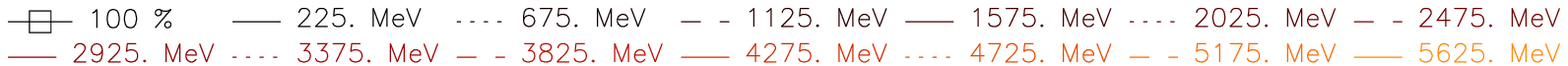}
\end{center}
\caption{Absolute values of the 
truncated spectral sums for the Polyakov loop correlator $C(d)$
for our $12^3 \times 6$ lattices below (lhs.\ plot) and above $T_c$ (rhs.). 
The symbols with error bars represent the full correlator as obtained 
directly from the gauge fields. The other curves are obtained from the spectral 
sums at various values of $\lambda_{cut}$, starting from a small cutoff 
for the curves at the bottom.}
\label{truncsums}
\end{figure}

In Fig.~\ref{truncsums} we show the absolute values of
the truncated sums for our ensembles on  $12^3
\times 6$ lattices below (lhs.~plot) and above $T_c$ (rhs.). The symbols
represent the data from an evaluation of the correlators $C(d)$ using directly 
the gauge links, and thus are the limiting values which the spectral sums
approach as the cutoff $\lambda_{cut}$ is sent towards the largest eigenvalue.
Only for those we show statistical errors, which are however representative also
for the truncated 
spectral sums. In the confined phase, the correlators show a cosh-type
of behavior in $d$  which is a consequence of a non-vanishing string tension for
the potential $V(d)$ in (\ref{loopcorr}) and the periodic boundary conditions
for the gauge fields\footnote{We remark, that the rather small lattices 
which we 
are restricted to by the need to compute the full Dirac spectrum lead to sizable
statistical errors towards the minimum of the cosh. The error bars appear
asymmetrically due to the logarithmic scale.}. For $T > T_c$ (rhs.~plot) the
symbols quickly form a constant plateau as is expected in the deconfined phase
where the string tension vanishes.        

The curves without symbols and error bars are the results from the spectral sums
at different values of the cutoff $\lambda_{cut}$, with the curves at the bottom
of the plots corresponding to a low $\lambda_{cut}$. We remark that due to the
different values of the gauge coupling which we use below and above $T_c$, a
different lattice cutoff results, such that for the ensemble below $T_c$
(lhs.~plot) the largest eigenvalues are near $\lambda_{max} \sim$
3500 MeV, while above $T_c$ (rhs.)
they reach up to $\lambda_{max} \sim$ 5800 MeV. 
Both phases show a similar behavior of a slow buildup
of the correlator as $\lambda_{cut}$ is increased. At about 30 \% of the maximal
eigenvalue  (e.g., the $\lambda_{cut} =$ 1125 MeV
curve for the lhs., and the $\lambda_{cut} =$ 1575 MeV curve on the
rhs.) the shape of the correlator, cosh-type and flat, respectively, 
is quite visible. As
$\lambda_{cut}$ is increased further, fluctuations appear which might be a
consequence of the fact that contributions from large eigenvalues are weighted
much stronger. Towards the UV end, the spectral sums even overshoot the final
result and settle at the exact values (symbols in the plot) only when all
eigenvalues are taken into account. This overshooting phenomenon was observed
already in \cite{paper2} where the expectation value of a single Polyakov loop
was studied. 

\vskip6mm
\noindent
{\Large Spatial structure of the eigenvector density correlators}
\vskip3mm
\noindent
Having established that the spectral sums give rise to the correct Polyakov loop
correlator and carry the information about confinement versus deconfinement also
when not fully summed, we now come to analyzing individual parts in the spectral
sums. In particular we discuss the role of the eigenvector densities
$\rho^{(k)}_i(\vec{x})$ in the two formulas (\ref{speccorr}) and
(\ref{speccorrexp}). 

The correlator (\ref{speccorrexp}) is related to the static
quark potential as stated in Eq.~(\ref{loopcorr}). Below $T_c$ the theory is
confining with a non-zero string tension $\sigma$, such that for
sufficiently large $|\vec{x} - \vec{y}|$ we expect $\langle L(\vec{x})
L^*(\vec{y}) \rangle \propto \exp(-N \sigma |\vec{x} - \vec{y}|)$. Above the
critical temperature the confinement is gone, such that $\langle L(\vec{x})
L^*(\vec{y}) \rangle$ is essentially constant for large $|\vec{x} - \vec{y}|$.
The characteristic spatial behavior of the correlator (\ref{speccorrexp}) below
and above $T_c$ can only come from the  eigenvector densities
$\rho^{(k)}_i(\vec{x})$, since their correlators are the only terms that contain
a dependence on the spatial coordinates $\vec{x}$ and $\vec{y}$. Consequently
the correlators of the eigenvector densities must already contain the spatial
information which in the full correlator (\ref{speccorrexp}) gives rise to the
discussed spatial behavior governed by the static potential 
below and above $T_c$.

In order to study the spatial structure of the (weighted) correlators
of the eigenvector densities, we analyze 
\begin{equation}
C^{(\lambda,\mu)}_{r,s}(d) \; = \; 
\frac{1}{3 L^3} \sum_{\vec{x}} \sum_{j = 1,2,3} \,
\Big\langle ( \, \lambda \, \mu \,)^N \,
\rho^{(\lambda)}_r(\vec{x}) \, 
\rho^{(\mu)}_s(\vec{x}+d \, \widehat{e}_j) \Big\rangle \; .
\label{rhocorrs}
\end{equation}
$\lambda$ and $\mu$ are eigenvalues on the positive imaginary axis and
$\rho^{(\lambda)}_r(\vec{x}), \rho^{(\mu)}_s(\vec{x}+d \, \widehat{e}_j)$ 
are the corresponding
eigenvector densities. 
The indices $r,s = 1,2,3$ refer to our three boundary conditions
$z_1,z_2, z_3$ as stated in (\ref{boundaryphase}). 

The correlators (\ref{rhocorrs}) are rather involved objects. Beside the
dependence on the distance $d$, they have four different parameters: two labels
$\lambda, \mu$ for the eigenvalues and two labels $r,s$ for the 3 different
boundary conditions. Thus we have to consider correlators where $\lambda = \mu$,
as well as correlators with $\lambda \neq \mu$. Furthermore our formulas
(\ref{speccorr}), (\ref{speccorrexp})  contain all possible combinations of
boundary conditions. In particular in the formula for the ensemble average
(\ref{speccorrexp}) the correlators with equal boundary conditions, $r=s$, come
with a plus sign, while the correlators with differing boundary conditions, $r
\neq s$, are subtracted. All these possible combinations of the parameters
$\lambda, \mu, r, s$, might give rise to a different dependence of the
corresponding correlators on the spatial distance $d$. In (\ref{speccorrexp})
these correlators are combined with relative signs and only this
combination then gives the correct dependence of the Polyakov loop correlators
on the spatial separation $d$. 

The central question we want to answer in this section is whether also the
individual eigenvector density correlators show a change in their spatial
behavior as one crosses from the confined into the deconfined phase. A
changing spatial spatial behavior of the eigenvector density correlators would
be the simplest mechanism for changing the static potential from confinement
to deconfinement.

While for the behavior of Dirac eigenvectors with small eigenvalues quite some
information is available, very little is known about Dirac eigenvectors with
eigenvalues that are not in the deep IR. Thus we here systematically study the
eigenvector density 
correlators (\ref{rhocorrs}) for various combinations of the parameters
$\lambda, \mu, r, s$. Since we here are predominantly interested in the decay
properties of these correlators, we normalize them such that they assume the
value 1 at vanishing distance $d=0$. 

\begin{figure}[t]
\begin{center}
\hspace*{-0mm}
\includegraphics[height=68mm,clip]{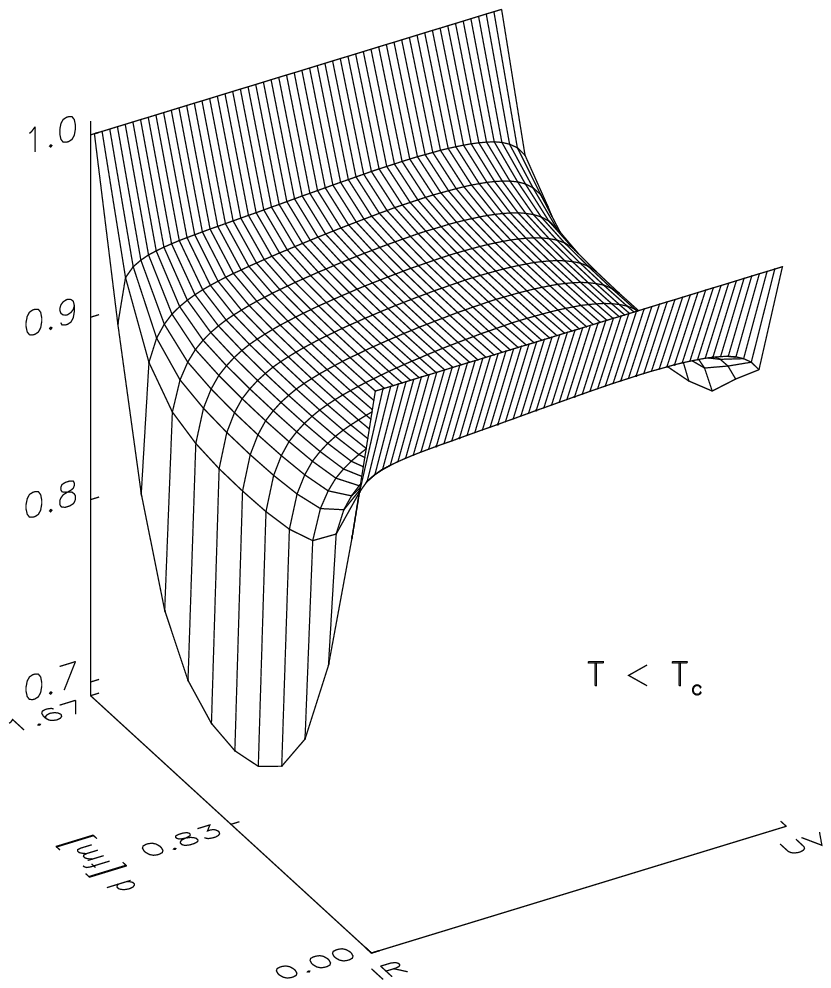}
\hspace{6mm}
\includegraphics[height=68mm,clip]{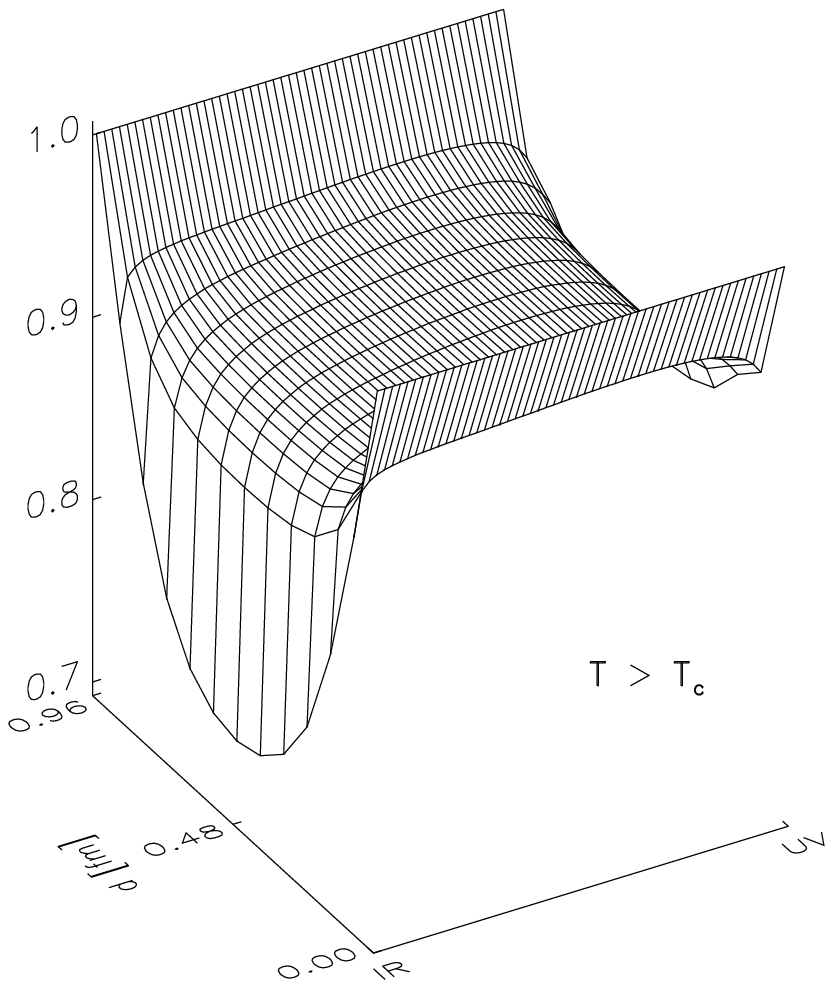}
\end{center}
\caption{
Normalized correlator $C^{(\lambda,\lambda)}_{r,r}(d)$ for equal eigenvalues 
as a function of $d$ and $\lambda$. The boundary conditions are chosen 
periodic, i.e., $r=1$. The lhs.~plot is for $T < T_c$, while the rhs.~is 
for the deconfined phase above $T_c$.}
\label{rhocorr1}
\end{figure}

In Fig.~\ref{rhocorr1} we begin with the simplest case, where we consider the
correlation $C^{(\lambda,\lambda)}_{r,r}(d)$ of an eigenvector with itself, i.e.,
both densities $\rho^{(\lambda)}_r(\vec{x})$ in the correlator correspond to the
same eigenvalue $\lambda = \mu$ and have the same boundary conditions which we
set to $r=s=1$, i.e., we consider periodic boundary conditions for the Dirac
operator. We remark, that the other two cases where the boundary conditions are
equal, $r=s=2$ and $r=s=3$, give similar results. The lhs.~plot shows the
correlator below $T_c$, while the rhs.~is for the deconfined phase $T>T_c$. On
the axis from the front to the back the spatial separation $d$ is plotted, while
on the axis running from left to right, we plot the number of the eigenvalue
$\lambda$, where at the left we start with the IR end of the spectrum and the
UV eigenvalues are at the right-hand side. 
Note that the eigenvalues are purely imaginary
and each eigenvalue comes with both signs. As we have already discussed it is
sufficient to sum over only one half of the eigenvalues, and in the
corresponding axis of the plot we simply run through the eigenvalues on 
the positive imaginary axis. 
These eigenvalues were divided into bins, and the correlators
averaged over the bins is what is shown in the 3-D mesh plot.  

For both cases, $T < T_c$ and $T > T_c$ the correlator decays with $d$, which
due to the periodicity gives rise to a minimum in the center followed by
another increase. The decay is most
prominent for the eigenvalues at the IR and UV ends of the spectrum, while near
the middle of the spectrum a rather flat trough is formed. A more detailed
analysis of slices through the 3-D plot reveals that at the IR and UV ends the
correlator indeed gives rise to a cosh, which shows that the
correlator decays exponentially. Slices through the center part of the plot show
that for the bulk of the spectrum the correlators of densities with medium sized
eigenvalues become flat.

However, the
most remarkable finding of the plots is the fact, that the normalized
density correlators
below and above $T_c$ are almost indistinguishable. In Fig.~\ref{rhocorr1} this
is only a finding for a very particular correlator, namely the one where the
eigenvalues of
the two densities and their boundary conditions coincide, but we will show,
that for essentially all normalized correlators contributing in
(\ref{speccorr}) and (\ref{speccorrexp}) the difference between the 
confined and deconfined phases is very small.

\begin{figure}[t]
\begin{center}
\includegraphics[height=48mm,clip]{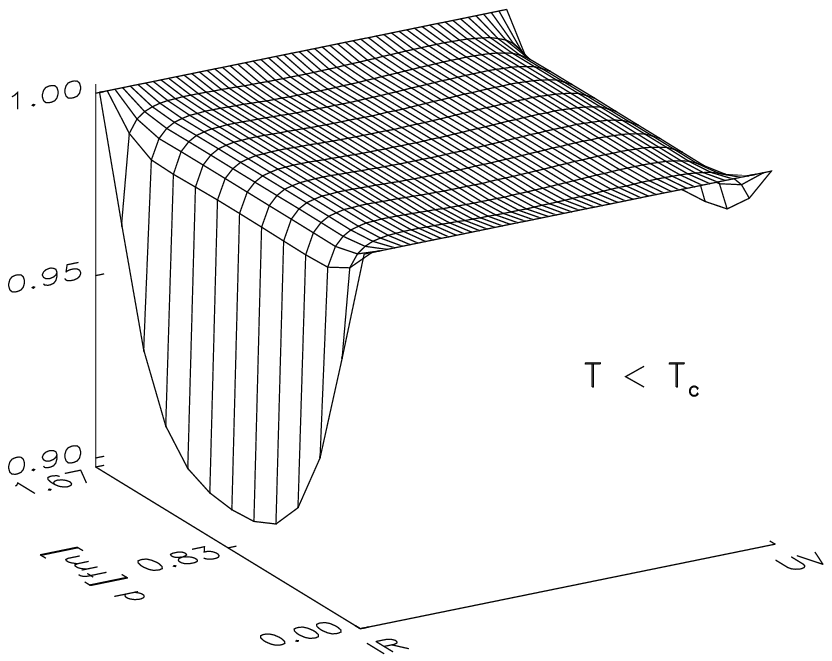}
\hspace{4mm}
\includegraphics[height=48mm,clip]{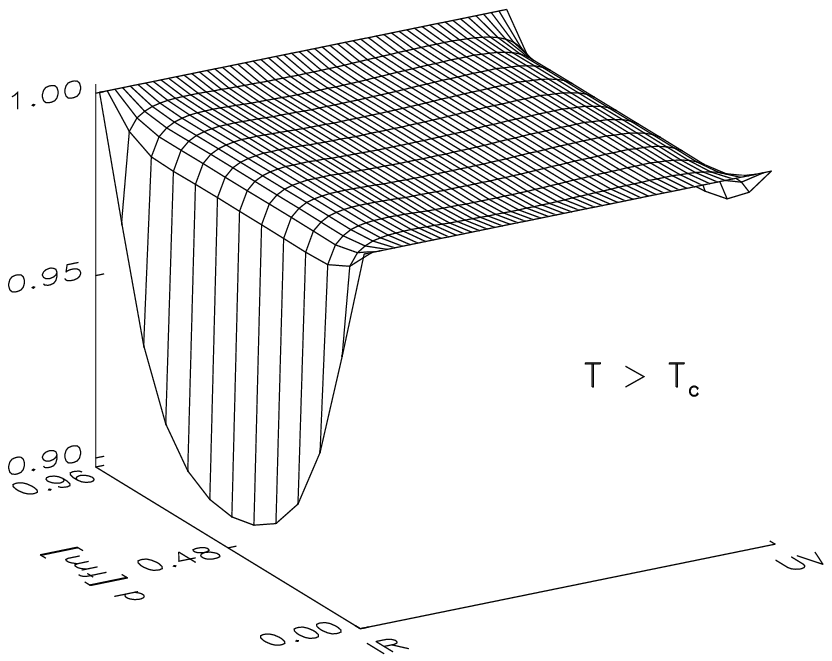}
\end{center}
\caption{
Same as Fig.~\ref{rhocorr1}, but now the correlator 
$C^{(\lambda,\lambda)}_{1,2}(d)$ which combines periodic 
conditions and the density with boundary phase $2\pi/3$ is shown.}
\label{rhocorr2}
\end{figure}

\begin{figure}[t]
\begin{center}
\includegraphics[height=46mm,clip]{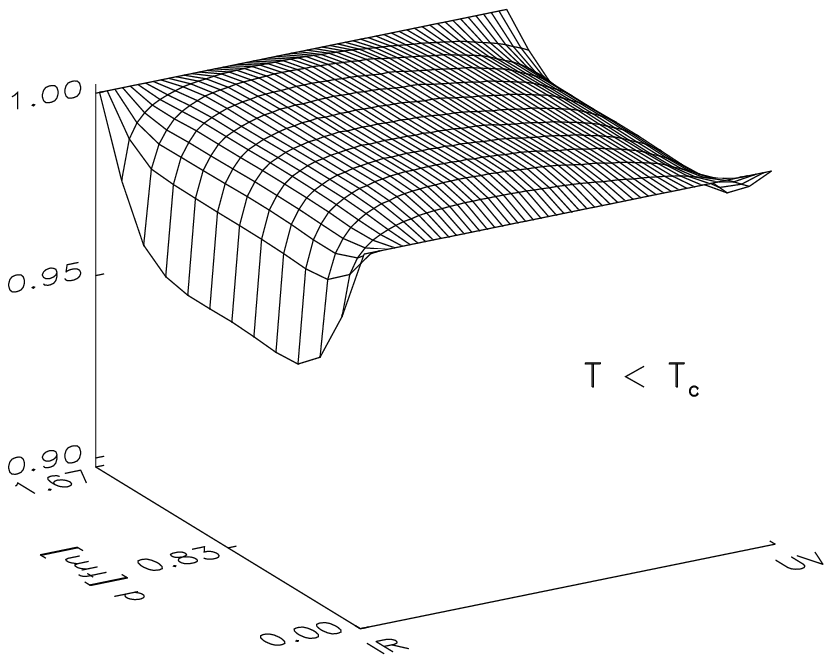}
\hspace{4mm}
\includegraphics[height=46mm,clip]{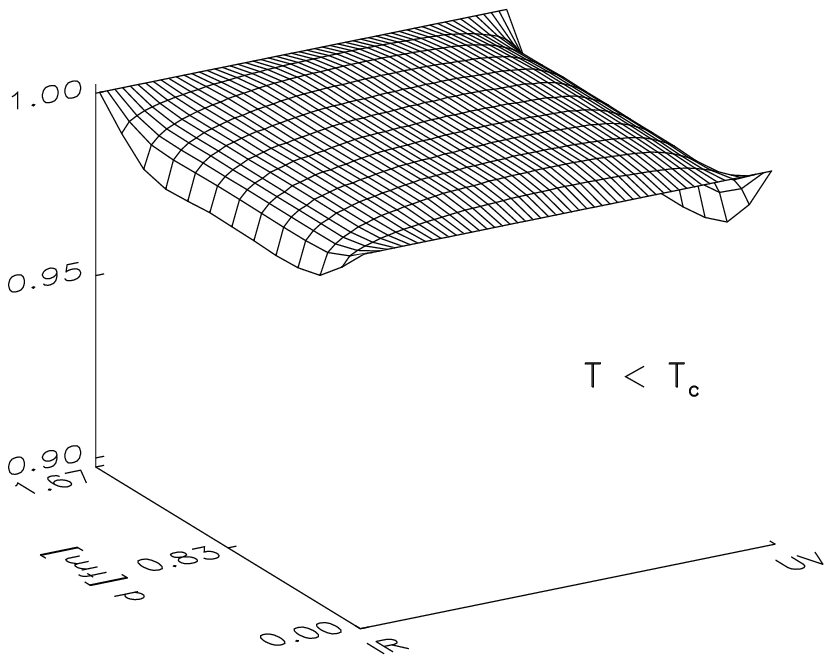}

\vskip6mm

\includegraphics[height=46mm,clip]{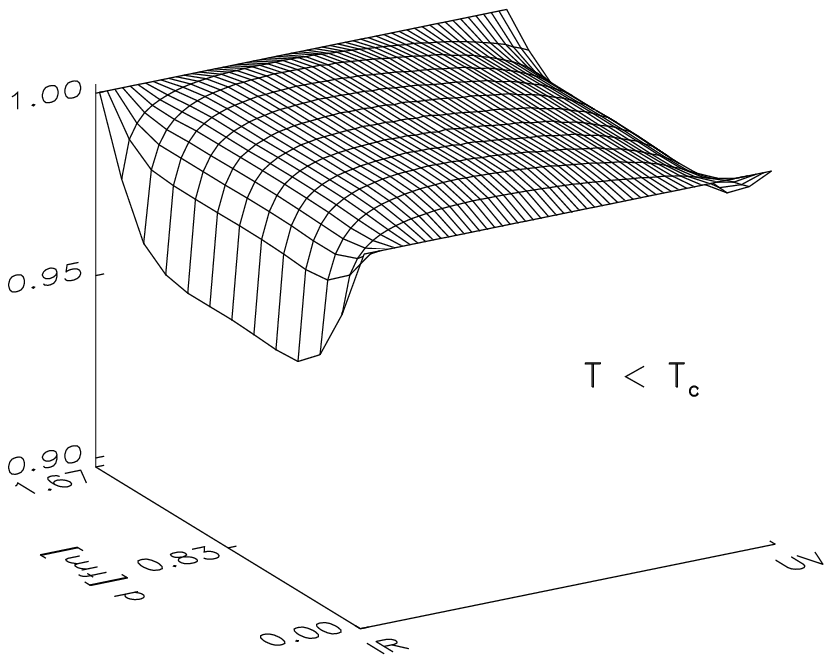}
\hspace{4mm}
\includegraphics[height=46mm,clip]{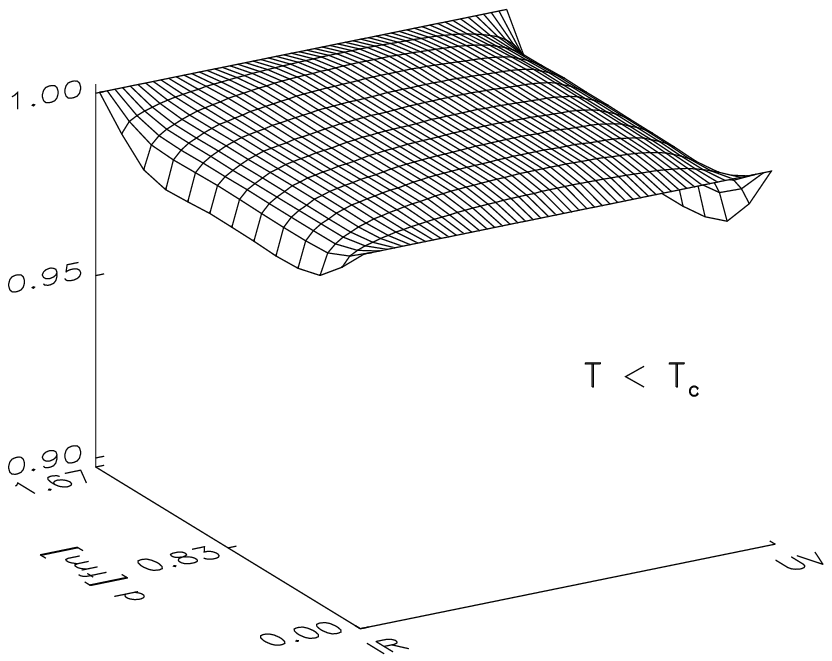}

\end{center}
\caption{
The normalized correlators 
$C^{(\lambda,\mu)}_{r,s}(d)$, where $\mu$ is held fixed at an IR
value (lhs.~column of plots) and at an UV value (rhs.). In the top row we show
the correlator where both densities are for periodic boundary conditions,
while in the bottom row, the combination of periodic with $2\pi/3$ boundary
conditions is displayed. All plots are for $T < T_c$.}
\label{rhocorr3}
\end{figure}

The next correlator of densities we study is the combination where the two 
densities still correspond to the same eigenvalue, but have different 
boundary conditions.  In particular we consider the combination of a periodic 
boundary condition and one with phase $2\pi/3$. The normalized correlator 
$C^{(\lambda,\lambda)}_{1,2}(d)$ for this choice of parameters is shown 
in Fig.~\ref{rhocorr2}. As in the case of equal boundary conditions 
we see that a pronounced decay is visible only for the smallest and 
the largest eigenvalues, while for the eigenvalues in the center of the
 distribution again a rather flat plateau is found. Also here the difference 
between the confined (lhs.~plot) and the deconfined phase (rhs.) 
is rather small.

So far we have restricted ourselves to correlators 
where the two densities were for equal eigenvalues. 
In Fig.~\ref{rhocorr3} we now show some cases of normalized
correlators of densities $C^{(\lambda,\mu)}_{r,s}(d)$
corresponding to different eigenvalues, $\lambda \neq \mu$. We always keep 
the eigenvalue $\mu$ fixed and plot the correlator as a 
function of the other eigenvalue $\lambda$ and the spatial 
separation $d$. For the lhs.~column $\mu$ is chosen near the IR end of the
spectrum, while on the rhs.~$\mu$ is at the UV end. In the top row both
densities are for periodic boundary conditions, while in the bottom plots
periodic, combined with $2\pi/3$ boundary conditions are used. All plots in
Fig.~\ref{rhocorr3} are for $T < T_c$. It is obvious, that these correlators
mainly give rise to a flat spatial behavior. 
Again the corresponding correlators for $T > T_c$ are almost identical. 

From our analysis of the spatial behavior of  
$C^{(\lambda,\mu)}_{r,s}(d)$ we conclude, that the eigenvector density
correlators (\ref{rhocorrs}) show a different spatial decay depending on
the chosen combination of the parameters $\lambda, \mu, r$ and $s$. For some
values we identify a clear exponential decay, which due to the periodicity is
manifest as a cosh, while for other choices of the 
parameters the correlators are essentially
flat. The final formula (\ref{speccorrexp}) for the Polyakov loop correlator
contains a sum over all parameter combinations. The 
individual correlators then combine their specific different spatial 
shapes such that either the exponential decay for the confined phase, or the
flat behavior of the deconfined phase emerges in the sum for the complete
Polyakov loop correlator. 

The most remarkable observation for the normalized eigenvector density 
correlators is, however, the 
fact that they are almost unchanged if the results for $T< T_c$ and $T > T_c$ 
are compared. We quantified this observation by analyzing the difference of
the correlators $C^{(\lambda,\mu)}_{r,s}(d)$ for our ensembles below and above
$T_c$. We found that the relative difference is below one percent
throughout. The largest discrepancies (about one percent) were found at the IR
and UV edges of the distributions. 

We repeated the same analysis for the modified correlators
\begin{equation}
\widetilde{C}^{(\lambda,\mu)}_{r,s}(d) \; = \; 
\frac{1}{3 L^3} \sum_{\vec{x}} \sum_{j = 1,2,3} \,
\Big\langle \rho^{(\lambda)}_r(\vec{x}) \, 
\rho^{(\mu)}_s(\vec{x}+d \, \widehat{e}_j) \Big\rangle \; ,
\label{rhocorrsmod}
\end{equation}
where the weight factor $(\lambda \mu)^N$ is omitted. Although these
eigenvector density correlators are not directly related to the Polyakov
loops, their behavior is of general interest: Hadronic correlation
functions may be expressed with the spectral theorem and various correlators
of Dirac eigenvectors will appear in such a representation. When crossing the
phase transition, the changing hadron content might leave a trace in the
correlators of the Dirac eigenvectors. In our analysis of the correlators 
$\widetilde{C}^{(\lambda,\mu)}_{r,s}(d)$ we found that their general behavior
is very similar to the one of the weighted correlators 
$C^{(\lambda,\mu)}_{r,s}(d)$ as shown in Figs.~2 -- 4. By analyzing as before the
difference of the correlators above and below $T_c$ we find again, that
crossing the phase transition leads to changes of the correlators which are 
only in the one percent range. 

\vskip6mm
\noindent
{\Large Amplitudes of the density correlators}
\vskip3mm
\noindent
In the last section we studied the spatial shape of our correlators 
(\ref{rhocorrs}) and thus plotted them such that they were normalized to 
1 at $d=0$. We now focus on the actual amplitudes of these correlators
defined as the correlator $C^{(\lambda,\mu)}_{r,s}(0)$ taken at distance $d = 0$. 
These amplitudes are important for understanding how the individual
correlators combine in the formula (\ref{speccorrexp}) to build up the
Polyakov loop correlation function. 

\begin{figure}[t]
\begin{center}
\includegraphics[height=48mm,clip]{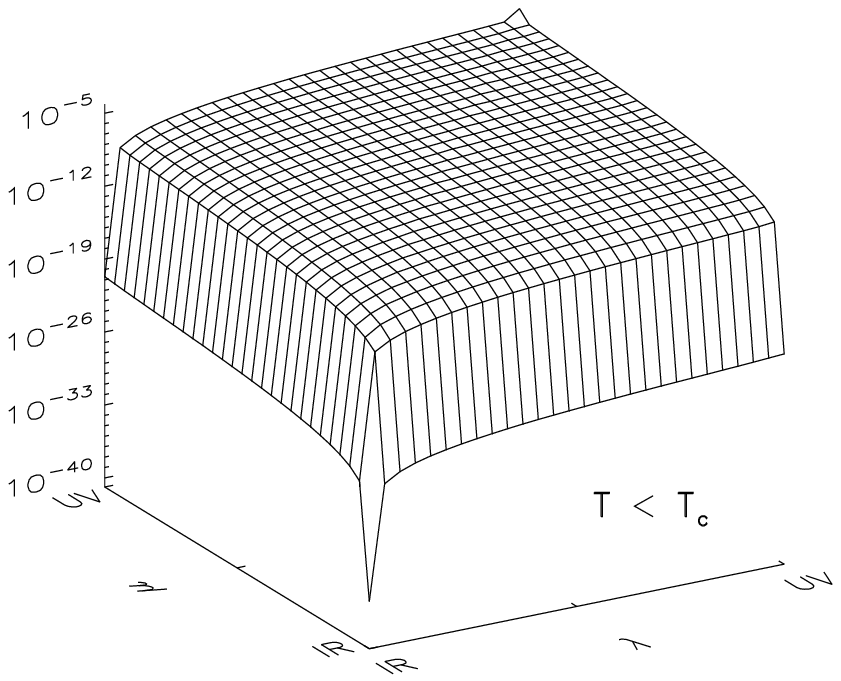}
\hspace{4mm}
\includegraphics[height=48mm,clip]{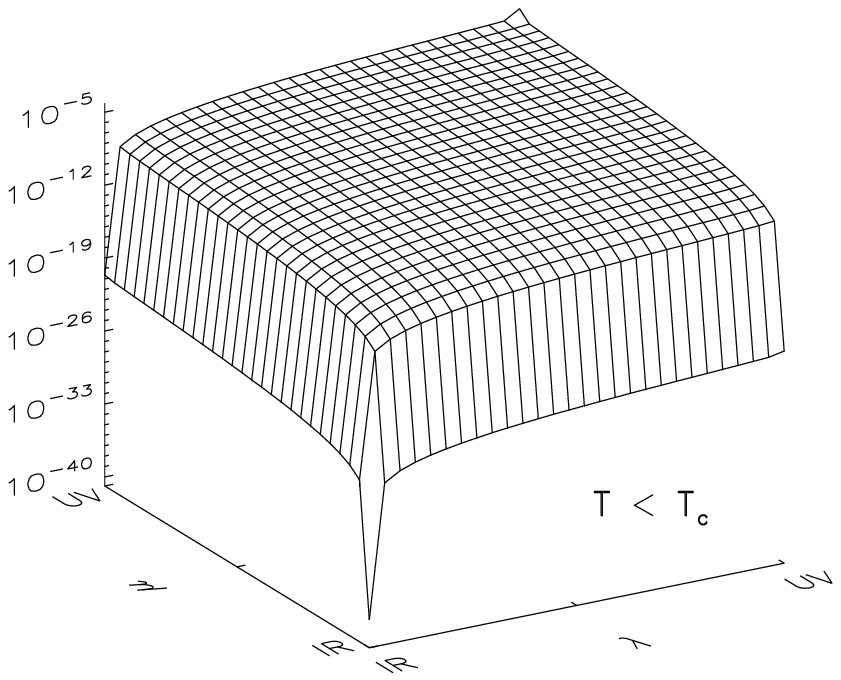}

\includegraphics[height=48mm,clip]{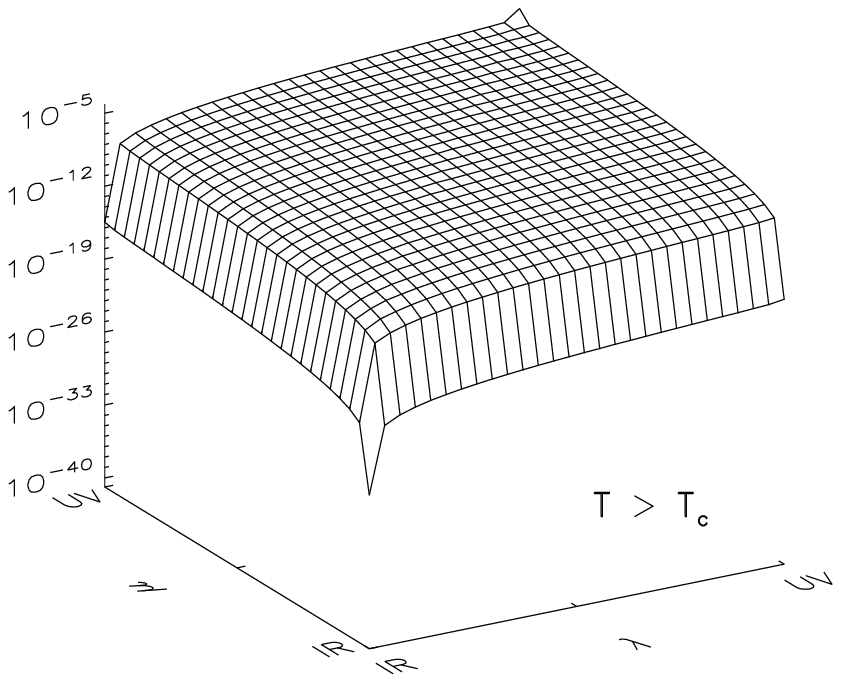}
\hspace{4mm}
\includegraphics[height=48mm,clip]{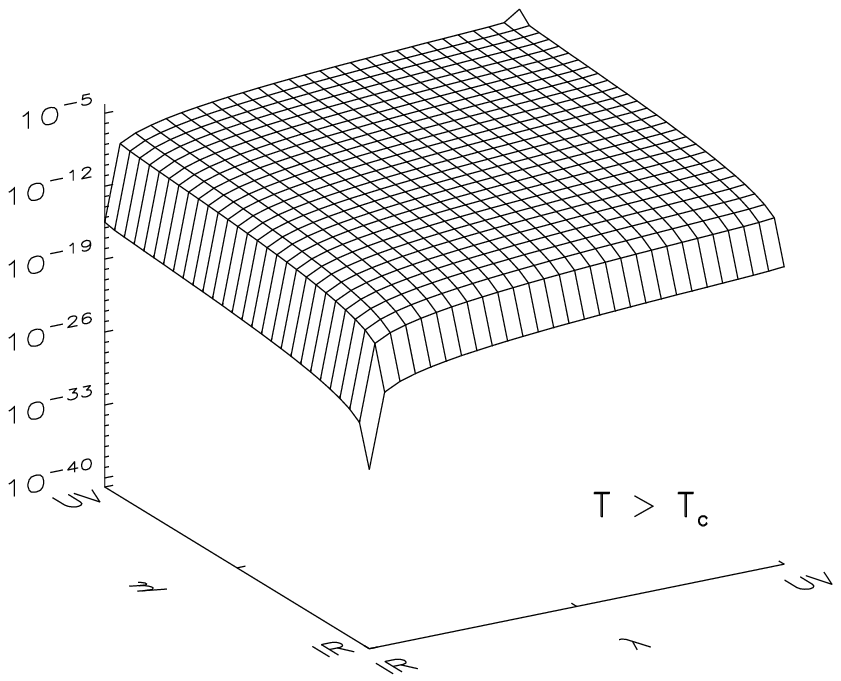}
\end{center}
\caption{
The amplitudes  
$C^{(\lambda,\mu)}_{r,s}(0)$ as a function of the eigenvalues $\lambda$ and
$\mu$. The plots on the lhs.~are for the boundary conditions $r=s=1$, 
while at the rhs.~we show $r=1, s=2$. In the top row we show the amplitudes for 
$T < T_c$, while at the bottom we have $T > T_c$.}
\label{amplitudes}
\end{figure}

In Fig.~\ref{amplitudes} we plot the amplitudes $C^{(\lambda,\mu)}_{r,s}(0)$
as a function of $\lambda$ and $\mu$ and compare different
combinations of the boundary condition parameters $r$ and $s$. The plots 
on the lhs.~are for boundary conditions $r=s=1$, and on the rhs.~the combination  
$r=1, s=2$ is shown. The top row is for $T < T_c$, while at the bottom $T >
T_c$ is shown. The plots show that the amplitudes for the contributions of 
different eigenvalues all have a very similar size, and only at the IR end 
show a sharp decrease. Furthermore the two combinations of boundary conditions 
($r=s=1$ in the lhs.~plot and $r=1, s=2$ on the rhs.) give rise to almost 
indistinguishable amplitude distributions. The same is true for the other 
combinations of boundary conditions that enter in the spectral sum 
(\ref{speccorrexp}). 

If we now compare the amplitudes below $T_c$ (top plots) to their counterparts 
above $T_c$ (bottom), we find a sizable discrepancy for some ranges of the 
eigenvalues $\lambda$ and $\mu$ (note that we use a logarithmic scale on the 
vertical axis). To illustrate this shift of the amplitudes more clearly we 
define the relative amplitudes
\begin{equation}
\Delta^{(\lambda,\mu)}_{r,s} \; = \; 
\frac{
C^{(\lambda,\mu)}_{r,s}(0) \Big|_{T<T_c} \; - \; \,
C^{(\lambda,\mu)}_{r,s}(0) \Big|_{T>T_{c_{\;}}}
}{
C^{(\lambda,\mu)}_{r,s}(0) \Big|_{T<T_c}  \; + \;  
C^{(\lambda,\mu)^{\;}}_{r,s}(0)\Big|_{T>T_c} 
} \; .
\label{reldef}
\end{equation}
The relative amplitudes $\Delta^{(\lambda,\mu)}_{r,s}$ may assume values 
between $-1$ and $1$ and are positive when the correlator amplitude is larger
in the confined phase, while it is negative otherwise. For identical
amplitudes one has $\Delta^{(\lambda,\mu)}_{r,s} = 0$.

\begin{figure}[t]
\begin{center}
\includegraphics[height=48mm,clip]{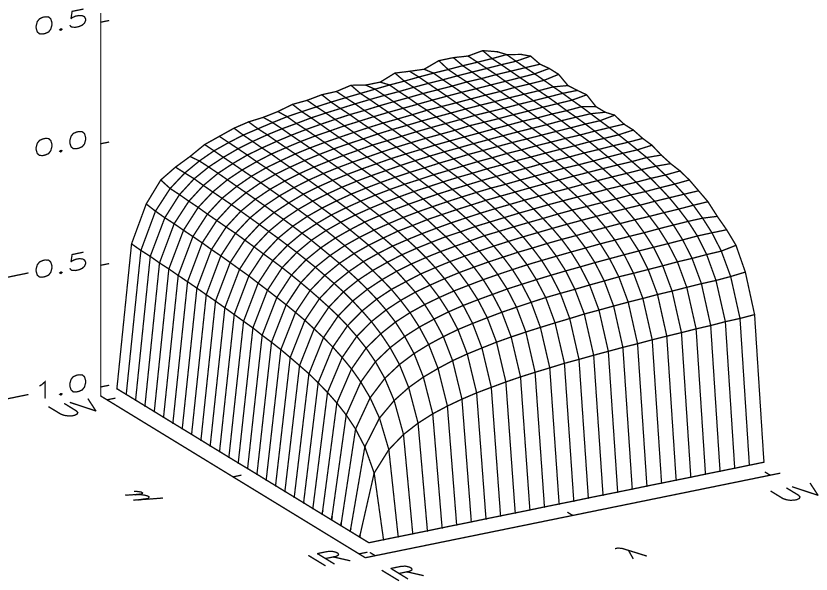}
\hspace{4mm}
\includegraphics[height=48mm,clip]{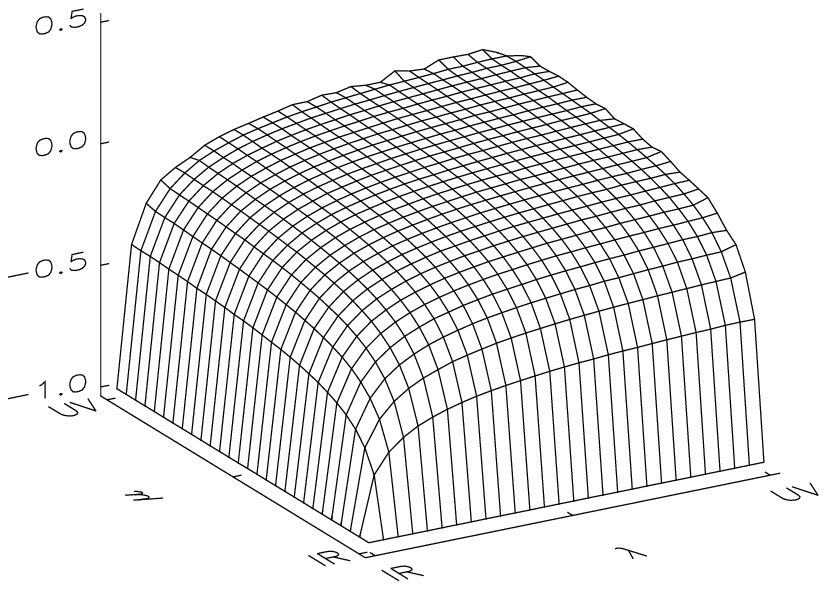}
\end{center}
\caption{
The relative amplitudes $\Delta^{(\lambda,\mu)}_{r,s}$ as 
defined in (\ref{reldef}) as a function of
$\lambda$ and $\mu$. The lhs.~plot is for boundary 
conditions $r=s=1$, the rhs.~for $r=1, s=2$. }
\label{relamplitudes}
\end{figure}

In Fig.~\ref{relamplitudes} we show the relative amplitudes 
$\Delta^{(\lambda,\mu)}_{1,1}$ on the lhs.~and $\Delta^{(\lambda,\mu)}_{1,2}$
on the rhs. The two different combinations of boundary conditions give rise 
to very similar behavior. Both relative amplitudes have in common that they 
are negative except when both eigenvalues $\lambda$ and $\mu$ are in the
upper third towards the UV end of the spectrum. This implies that the amplitudes 
in the IR are smaller in the confined phase, while towards the UV part of 
the spectrum the amplitudes are larger in that phase. Thus in the confined 
phase more weight is on the UV end of the spectrum, while in the deconfined 
phase the IR part is enhanced. This shift of the amplitudes changes the 
weight of eigenvector density correlators with different decay properties, 
which we analyzed in the last section, and in this way switches from the 
confining to the deconfining static potential. Since the spatial decay 
properties of the eigenvector density correlators
are essentially invariant, it is this relative shift of the 
amplitudes which we identify as the mechanism giving rise to the 
changing behavior at $T_c$. 

\newpage
\noindent
{\Large Conclusions}
\vskip3mm
\noindent
In this article we have studied correlators of Polyakov loops and their 
spectral decomposition in terms of eigenvalues and eigenvectors of the
Dirac operator.  
The spectral sums we discuss connect the static quark-antiquark potential 
to correlators of densities of the Dirac eigenvectors. From those eigenvector 
density correlators the static potential must build up its spatial dependence, 
characteristic for the confined and deconfined phases. The eigenvector density
correlators were studied numerically for quenched SU(3) configurations below
and above the transition temperature $T_c$. 

We find that depending on the parameters (eigenvalue combination and boundary
conditions), the individual eigenvector density correlators have different decay
properties, ranging from a strong exponential decay to a rather flat
behavior. The final Polyakov loop
correlator is obtained as a combination of all these contributions and only 
for the full spectral sum the spatial decay properties
of the Polyakov loop correlator emerge.    

An interesting outcome of the numerical part of our analysis 
is the finding that the spatial decay properties of the (normalized) 
correlators of the eigenvector densities are essentially the same in 
the confined and the deconfined phases. The relative difference of the normalized 
correlators below and above $T_c$ was found to be less than one percent. 
For the amplitudes the situation is different. We show that they change when 
crossing $T_c$, such that in the deconfined phase the IR part is enhanced in 
the spectral sums, while the UV part is reduced. This shift of the amplitudes 
leads to a different combination of spatial decay properties of individual 
correlators and the static potential switches from confinement to deconfinement.

Having established a particular behavior of Dirac eigenvector correlators 
in the two phases as mechanism for shifting from confinement to deconfinement, 
a more systematic analysis of general eigenvector correlators appears 
interesting. Since with the spectral theorem hadronic observables may be 
expressed in terms of Dirac eigenvector correlators, a better understanding 
of such correlators might lead to a clearer characterization of hadronic 
properties in the deconfined phase.

\vskip6mm
\noindent
{\Large Acknowledgments}
\vskip3mm
\noindent
We thank Falk Bruckmann, Christian Hagen, Christian Lang, 
Kurt Langfeld, Wolfgang S\"oldner, Pierre van Baal, Jac Verbaarschot and 
Andreas Wipf for discussions. The numerical analysis was
done at the ZID, University of Graz. E.~Bilgici
is supported by the FWF Doktoratskolleg {\sl Hadrons in Vacuum, Nuclei and
Stars} (DK W1203-N08) and C.~Gattringer by FWF Project 20330.


\newpage
\begin{thebibliography}{1234567}

\vspace{-1mm}


\bibitem{paper1}
  C.~Gattringer,
  Phys.\ Rev.\ Lett.\  {\bf 97}, 032003 (2006).

\bibitem{paper2}
  F.~Bruckmann, C.~Gattringer, C.~Hagen,
  Phys.\ Lett.\  B {\bf 647}, 56 (2007).

\bibitem{wipf}
  F.~Synatschke, A.~Wipf, C.~Wozar,
  Phys.\ Rev.\  D {\bf 75}, 114003 (2007).

\bibitem{paper3}
  C.~Hagen, F.~Bruckmann, E.~Bilgici, C.~Gattringer,
  PoS(LATTICE 2007)289 [arXiv:0710.0294 [hep-lat]].
  
\bibitem{soeldner}
W.~S\"oldner, PoS(LATTICE2007) 222.

\bibitem{paper4}
  E.Bilgici, F.~Bruckmann, C.~Gattringer, C.~Hagen, 
  arXiv:0801.4051 [hep-lat].

\bibitem{wipf2}
  F.~Synatschke, A.~Wipf and K.~Langfeld,
  arXiv:0803.0271 [hep-lat].


\bibitem{LuWe}
  M.~L{\"u}scher, P.~Weisz, Commun.~Math.~Phys.\ {\bf 97}, 59 (1985); 
  E: {\bf 98}, 433 (1985); 
  G.~Curci, P.~Menotti, G.~Paffuti, Phys.~Lett.~B {\bf 130}, 205 (1983); 
  E: B {\bf 135}, 516 (1984).

\bibitem{scale}
  C.~Gattringer, R.~Hoffmann, S.~Schaefer,
  Phys.\ Rev.\ D {\bf 65}, 094503 (2002).

\end{thebibliography}
\end{document}